%% file: main.tex
\documentclass[runningheads,orivec]{llncs}
\usepackage[utf8]{inputenc}
\usepackage[T1]{fontenc}

\input{preamble}

\begin{document}
\title{Toward Quantum Utility in Finance: A Robust Data-Driven Algorithm for Asset Clustering}

\titlerunning{A Robust Quantum Algorithm for Asset Clustering}

\author{Shivam Sharma\inst{1}\orcidID{0009-0001-6580-1058} \and
Supreeth Mysore Venkatesh\inst{2,3}\orcidID{0000-0002-9824-7399}
\and
Pushkin Kachroo\inst{1}\orcidID{0000-0002-7701-8925}}
\authorrunning{S. Sharma et al.}

\institute{University of Nevada, Las Vegas, Nevada, USA \\
\email{sharms15@unlv.nevada.edu, pushkin@unlv.edu} \and
University of Kaiserslautern-Landau (RPTU), Kaiserslautern, Germany \email{supreeth.mysore@rptu.de} \and
German Research Center for Artificial Intelligence (DFKI), Saarbruecken, Germany \email{supreeth.mysore@dfki.de}}

\maketitle

\begin{abstract}
Clustering financial assets based on return correlations is a fundamental task in portfolio optimization and statistical arbitrage. However, classical clustering methods often fall short when dealing with signed correlation structures, typically requiring lossy transformations and heuristic assumptions such as a fixed number of clusters. In this work, we apply the Graph-based Coalition Structure Generation algorithm (GCS-Q) to directly cluster signed, weighted graphs without relying on such transformations. GCS-Q formulates each partitioning step as a QUBO problem, enabling it to leverage quantum annealing for efficient exploration of exponentially large solution spaces. We validate our approach on both synthetic and real-world financial data, benchmarking against state-of-the-art classical algorithms such as SPONGE and $k$-Medoids. Our experiments demonstrate that GCS-Q consistently achieves higher clustering quality, as measured by Adjusted Rand Index and structural balance penalties, while dynamically determining the number of clusters. These results highlight the practical utility of near-term quantum computing for graph-based unsupervised learning in financial applications.
\blfootnote{This work has been accepted for publication in the proceedings of the 1st International Quantum Engineering conference and exhibition (QUEST-IS 2025), held 1--4 December 2025, Paris, France.}

\end{abstract}

\section{Introduction and Background}
\label{sec:Background}

Information about relationships among financial assets has long been central to strategies in portfolio optimization and algorithmic trading. Since companies and their stock prices are interdependent, exploiting these correlations, especially in markets that are not perfectly efficient, can enhance decision-making in portfolio construction. The foundational work of Markowitz \cite{markowitz_portfolio} introduced the mean-variance optimization framework, establishing diversification as a means to maximize return while minimizing risk \cite{diversification}.

More recent strategies, such as statistical arbitrage \cite{BERTRAM20102234} and pair trading \cite{alvaro_cointegrated_assets}, use correlation or cointegration metrics but have become less effective due to increased market adoption. Consequently, graph-theoretic methods have emerged as robust alternatives \cite{statistical_arbitrage}, where assets are represented as nodes and correlation matrices are treated as signed, weighted graphs. Clustering these graphs aims to group assets with high intra-cluster and low inter-cluster correlations \cite{LEON20171334, TOLA2008235}, recent works focusing on identifying the best-performing clustering algorithms \cite{alvaro_correlation_clustering}.

Traditional clustering approaches such as $k$-Means, $k$-Medoids \cite{lloyds_kmeans, kaufman1990pam}, or hierarchical clustering \cite{hierarchical} face limitations when applied to signed graphs. These algorithms require positive-definite distance metrics and often rely on transforming signed correlation matrices $\rho \in [-1, 1]$ into non-negative distances using monotonic mappings like \cite{Mantegna1999, correlation_to_distance, CBIC2021-49}:
\begin{equation}\label{eq:distance_transform}
    d_{ij}^{(\alpha)} = \sqrt{\alpha(1 - \rho_{ij})}, \quad \alpha > 0
\end{equation}
Although this transformation preserves ranking, it loses semantic fidelity, for example, $\rho_{ij} = 0$ is mapped to $\sqrt{\alpha}$, incorrectly implying a fixed non-zero dissimilarity. Thus, even optimally solving the transformed problem does not recover the optimal clustering under the original signed graph.
Centroid-based methods further deviate from the true objective, which is to maximize intra-cluster correlations and minimize inter-cluster correlations, not to minimize distances to centroids. Moreover, these classical methods require manual tuning of hyperparameters such as the number of clusters $k$, initialization strategies, or threshold values for merging/dividing. These choices are not generalizable across datasets and must be revalidated for unseen data.
Spectral clustering methods provide a more natural fit for graph-structured data by using eigenvalue decomposition of the similarity matrix \cite{spectral_clustering}. SPONGE \cite{sponge}, a spectral method for signed graphs, has shown strong performance but is effective mainly when $k$ is large or the graph is very sparse \cite{alvaro_correlation_clustering, sponge_application}, conditions not typically met in financial data.

Quantum methods have also been explored, notably in portfolio optimization using gate-based variational circuits \cite{Buonaiuto2023}, though existing hardware limit scalability. One prior attempt at quantum-assisted clustering formulates the problem as a maximum clique and translates it into a large QUBO instance \cite{qfinance}, introducing additional binary variables and exceeding the capabilities of current annealers.

In this work, we propose applying the Graph-based Coalition Structure Generation algorithm (GCS-Q) \cite{gcsq} to the task of asset clustering using return correlation data. GCS-Q operates directly on the signed graph, avoiding lossy transformations. It starts with all assets grouped together and iteratively partitions the graph via minimum cuts, maximizing intra-cluster weights at each step and dynamically determining $k$.
This approach is well-suited to the signed graph setting where classical solvers offer only heuristic workarounds. The minimum-cut subproblem is formulated as a QUBO and solved on D-Wave's quantum annealer, leveraging its ability to explore exponentially large solution spaces efficiently.
We validate GCS-Q on both synthetic and real-world financial datasets, demonstrating its superior clustering quality compared to classical state-of-the-art methods.

\section{Methodology}

We consider a set of financial assets \( \mathcal{A} = \{a_1, a_2, \dots, a_n\} \), whose historical returns over a rolling window are denoted by \( r_i(t) \) for asset \( a_i \) at time \( t \). The pairwise Pearson correlation coefficients \cite{pearson_vii_1895} are computed as:
\begin{equation}
    \rho_{ij} = \frac{\text{Cov}(r_i, r_j)}{\sigma_{r_i} \sigma_{r_j}}, \quad \rho_{ij} \in [-1, 1].
\end{equation}

These correlations define a signed, weighted, undirected graph \( G = (V, E, w) \), where each vertex \( v_i \in V \) uniquely corresponds to asset \( a_i \), and each edge \( (v_i, v_j) \in E \) is weighted by \( w_{ij} = \rho_{ij} \). In contrast to traditional clustering methods that transform correlations into non-negative distances \cite{CBIC2021-49}, we retain the signed correlations, thereby preserving both positive (co-movement) and negative (anti-correlation) relationships.

We apply the GCS-Q algorithm, previously developed for coalition structure generation in graphs \cite{gcsq}, to this asset clustering setting. The core contribution lies in adapting GCS-Q to leverage its ability to operate directly on the signed graphs without resorting to heuristic transformations. 

Formally, GCS-Q aims to find a coalition structure \( \Pi = \{C_1, C_2, \dots, C_k\} \), a partition of \( V \) into disjoint subsets, that maximizes the intra-cluster edge weights:
\begin{equation}
    \label{eq:gcsq_objective}
    \max_{\Pi} \sum_{C \in \Pi} \sum_{\substack{i,j \in C \\ i < j}} w_{ij}.
\end{equation}
This objective promotes grouping of assets with strong positive correlations while implicitly disincentivizing inclusion of negatively correlated pairs within the same cluster. 

The GCS-Q algorithm proceeds iteratively, starting with the full graph \( G_0 = G \). At each iteration \( t \), it computes a minimum cut \( (S_t, \bar{S}_t) \) on the current subgraph \( G_{t-1} \), using the actual edge weights \( w_{ij} \) which is strictly NP-Hard. The resulting partition optimizes the objective in Eq.~\eqref{eq:gcsq_objective}. The process is then recursively applied to the subgraphs induced by \( S_t \) and \( \bar{S}_t \), eliminating the cut edges at each step, and terminates dividing a subgraph further when there is no cut whose value is lower than the sum of all the edges in that subgraph. This stopping criterion inherently infers the number of clusters \( k \) as part of the optimization, obviating the need for manual tuning or initialization required by classical approaches.

Each minimum cut in GCS-Q is formulated as a Quadratic Unconstrained Binary Optimization (QUBO) problem \cite{glover_qubo_tutorial}, well-suited for near-term quantum devices. These QUBOs can be efficiently solved using quantum annealers (e.g., D-Wave) or variational algorithms like the Quantum Approximate Optimization Algorithm (QAOA) \cite{quacs}, enabling hybrid quantum-classical workflows.

Theoretical work such as \cite{Bansal2004} has shown that optimally clustering signed graphs with edge weights in \( [-1, +1] \) is NP-Hard. Our practical contribution lies in demonstrating that GCS-Q can robustly address this challenge in financial asset clustering, highlighting its potential for real-world deployment and showcasing practical applicability of near-term quantum technologies, particularly quantum annealing.

\section{Experiments}

We conduct two sets of experiments. First, we benchmark the clustering quality of GCS-Q against classical baselines on synthetically generated signed graphs designed to mimic financial correlation structures. Second, we evaluate the methodology on real-world asset return data to demonstrate its practical utility.

All implementations are in \texttt{Python 3.12}. The QUBO subproblems in GCS-Q are solved using the D-Wave \texttt{Advantage\_system5.4} annealer, which comprises 5614 physical qubits and $40,050$ couplers arranged in the pegasus topology. The device is accessed remotely via the \texttt{dwave-ocean-sdk} APIs. As noted in \cite{q_seg}, most runtime overhead arises from internet latency and queue wait times due to the lack of onsite quantum hardware. As such, the experiments emphasize solution quality to assess the practical relevance of the methodology using exisitng quantum annealers.

All remaining computations are executed on a standard 64-bit workstation equipped with a 12th Gen Intel Core i7-12800H CPU and 64 GB RAM.

\subsection{Synthetic Data} \label{sec:synthetic-data}

We generate structurally balanced signed graphs, sampling intra-cluster edges from $[0.1, 1.0]$ and inter-cluster edges from $[-1.0, -0.1]$ to emulate realistic correlation patterns. Unlike sparse graphs in \cite{sponge} that use discrete weights $\in \{-1, 0, +1\}$ and equal cluster sizes, our graphs span a continuous range and incorporate non-uniform cluster sizes via Dirichlet-based stochastic allocation. This introduces heterogeneity while ensuring all clusters remain non-empty. No noise is added, isolating algorithmic behavior under ideal structural conditions. We extrapolate the setup later in Sec.~\ref{sec:yfinance-data} to evaluate robustness under noise.

We benchmark GCS-Q against classical baselines commonly used for clustering assets via return correlation data. SPONGE and its symmetric variant SPONGE\textsubscript{sym} \cite{sponge} are among the strongest graph-based techniques \cite{alvaro_correlation_clustering}. We also include \textit{k}-Medoids (via the PAM algorithm) as a robust baseline well-suited to financial data \cite{CBIC2021-49}.

Unlike GCS-Q, which determines the number of clusters $k$ dynamically, SPONGE and PAM require $k$ as input. This requirement often demands heuristics like silhouette analysis \cite{silhouette} or the elbow method \cite{elbow_method}, which are expensive and lack generalizability. Since SPONGE parallels spectral clustering, we estimate $k$ using the eigengap heuristic \cite{spectral_gap_eigen_heuristics}, which selects $k$ based on the largest gap between successive Laplacian eigenvalues.

\begin{figure}[H]
    \centering
    \includegraphics[width=\linewidth]{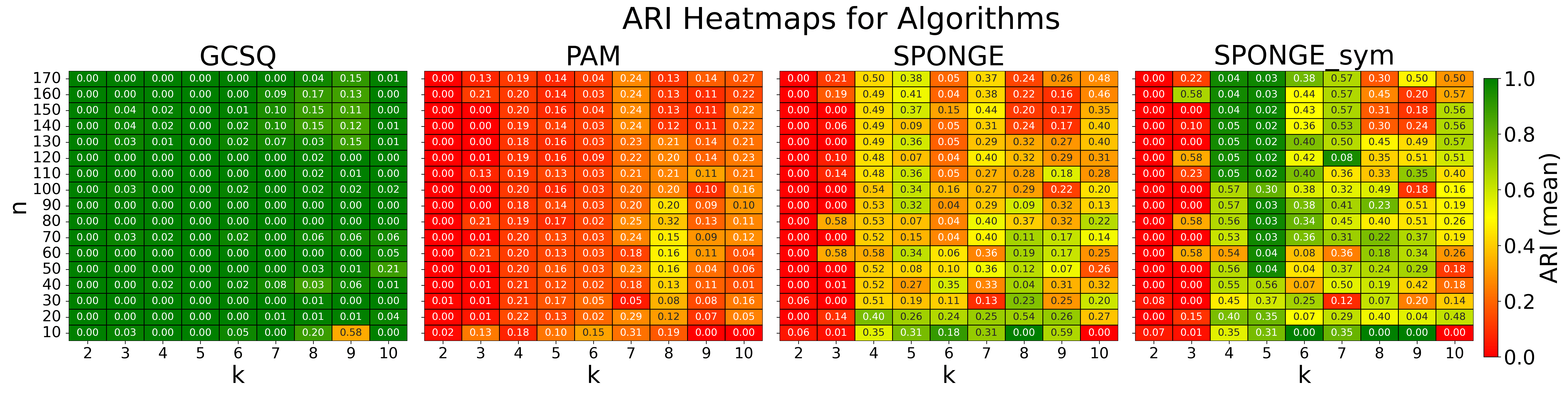}
    \caption{ARI scores for synthetic data of varying number of nodes $n$ and number of clusters $k$. Color represents mean and the (tiny) numerical annotations indicate variances across seeds.}
    \label{fig:ari}
\end{figure}

We conduct experiments varying total nodes $\in \{10, 20, \dots, 100\}$, with $k$ ranging from 2 to 10 across 33 random seeds. Since ground-truth clusters are known, we assess accuracy using Adjusted Rand Index (ARI) \cite{adjusted_rand_index}.

Although GCS-Q is limited to roughly 175 nodes due to qubit constraints and hardware topology, Fig.~\ref{fig:ari} shows it consistently outperforms classical solvers. PAM performs worst, largely due to its reliance on transforming correlations into non-negative distances. While SPONGE and SPONGE\textsubscript{sym} perform well on graphs with weights in $\{-1, 0, 1\}$ and balanced cluster sizes, their performance degrades on continuous weights and stochastic sizes.

Moreover, SPONGE is designed for sparse graphs \cite{alvaro_correlation_clustering}, which is misaligned with financial data where correlations close to zero are rare. The superior performance of GCS-Q is attributed to the expressive solution space explored by the quantum annealer at each step.

\subsection{Yahoo Finance Data} \label{sec:yfinance-data}

We collect hourly closing prices of 50 assets from diverse sectors from \texttt{yfinance}, compute rolling log returns, and derive the Pearson correlation matrix, which is interpreted as a signed adjacency matrix.
The same clustering solvers from Sec.~\ref{sec:synthetic-data} are applied. In the absence of ground truth, we assess clustering quality using the Penalty metric, which quantifies deviation from structural balance by penalizing negative intra-cluster correlations and positive inter-cluster correlations:
\[
\text{Penalty}(\Pi) = \sum_{C \in \Pi} \sum_{\substack{i,j \in C \\ i < j \\ w_{ij} < 0}} |w_{ij}| + \sum_{\substack{C_a \ne C_b}} \sum_{\substack{i \in C_a, j \in C_b \\ i < j \\ w_{ij} > 0}} w_{ij}
\]
A lower penalty indicates better clustering, with zero representing a perfectly balanced structure. Such structurally sound clusters enable more effective asset diversification, aligning with the goal of minimizing portfolio variance in the Markowitz framework \cite{markowitz_portfolio}.

\begin{figure}[ht]
    \centering
    \includegraphics[width=0.8\linewidth]{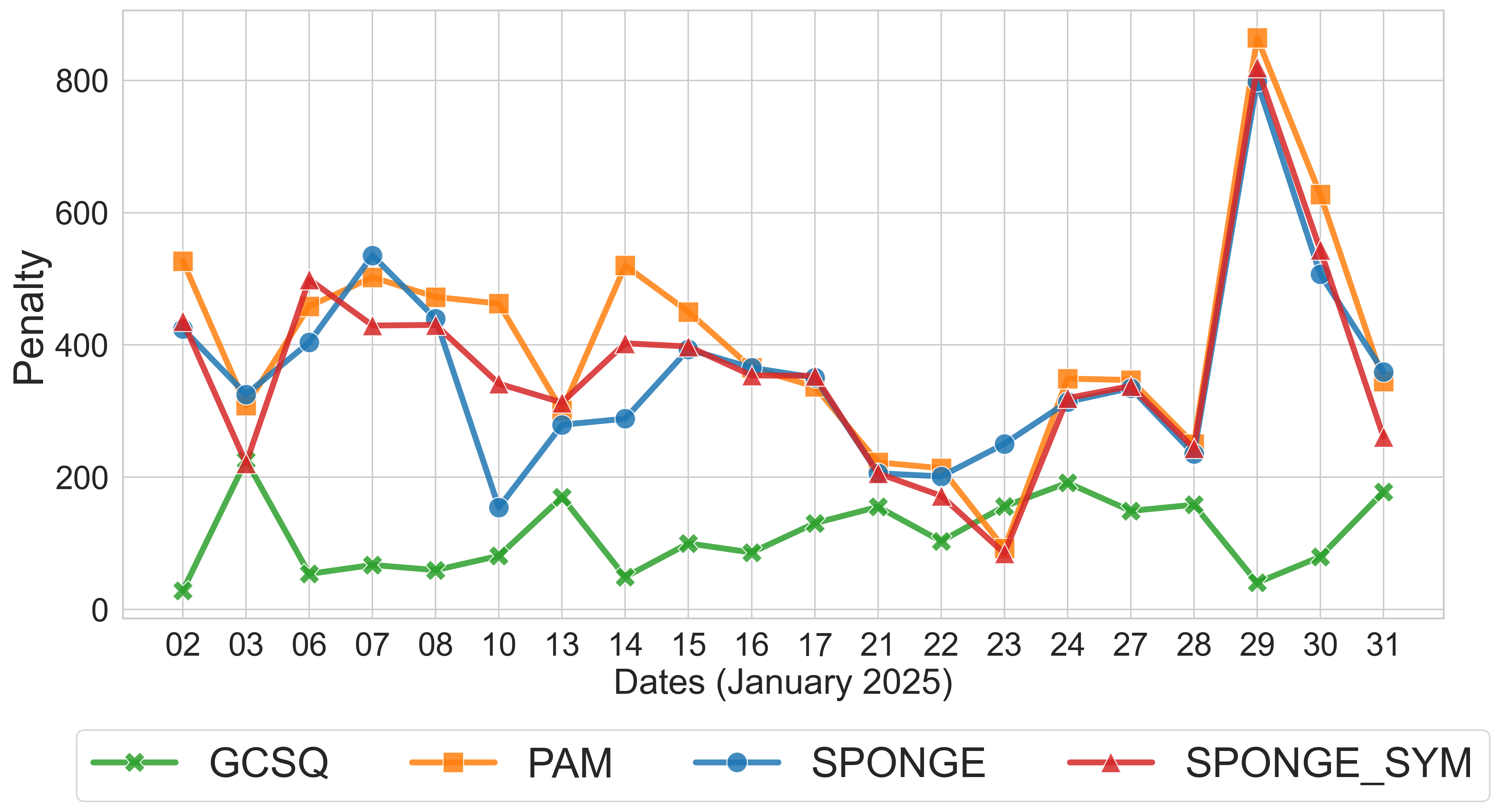}
    \vspace{-12pt}
    \caption{Penalty comparison across clustering algorithms over business days in January 2025, based on correlations derived from hourly return intervals.}
    \label{fig:penalties}
    \vspace{-12pt}
\end{figure}

As Fig.~\ref{fig:penalties} illustrates, GCS-Q consistently achieves the lowest penalty, indicating its superiority in uncovering market structures conducive to portfolio optimization and statistical arbitrage. 
Low penalty values indicate clusters that are internally cohesive, which supports mean-reversion strategies \cite{alvaro_correlation_clustering}, and externally distinct, making them well-suited for capturing diverse risk-return profiles essential for effective diversification \cite{diversification}.

\section{Discussion and Conclusion}

This work is the first, to our knowledge, to demonstrate a quantum advantage in solution quality using real hardware for the practical task of asset clustering, validated on both synthetic and real financial data. GCS-Q's strength lies in its exponential solution space exploration at each iteration. When executed classically, its complexity is $\mathcal{O}(n2^n)$, where $\mathcal{O}(2^n)$ stems from solving the minimum cut and $\mathcal{O}(n)$ from dynamically determining $k \in [1,n]$. While commercial solvers like Gurobi and CPLEX exist for combinatorial optimization, they struggle to scale on dense graphs \cite{leo_satellites}.
Although the adiabatic quantum process theoretically runs in constant time, apart from the hardware noise, current access via shared cloud infrastructure introduces latency and queue delays. In practice, clustering $n=170$ assets with $k=10$ took over 10 minutes on average, despite per-QUBO solve times being capped at 1 second.

GCS-Q is robust to real-world scenarios with heterogeneous cluster sizes and avoids the idealized constraints (e.g., uniform clusters, ternary edge weights) assumed in prior work \cite{sponge, alvaro_correlation_clustering}. It requires no manual tuning of $k$, operating in a fully data-driven manner. Unlike classical methods that convert signed correlations to Euclidean distances \cite{Mantegna1999, CBIC2021-49}, GCS-Q works directly on the signed graph, preserving relational structure.

Future directions include applying the resulting clusters to downstream tasks like asset allocation and evaluating performance using metrics such as Sharpe Ratio and returns. From a quantum perspective, deploying GCS-Q on D-Wave's hardware with Zephyr topology and better connectivity, exploring custom embeddings, and using gate-model quantum solvers with qubit-efficient strategies \cite{qubit_efficient_encoding}, as well as benchmarking against classical optimizers like Gurobi, are promising next steps. Moreover, similar iterative approaches are also used in optimizing machine learning models \cite{iqls}. This work, we believe, marks a concrete advancement toward near-term quantum utility for signed graph clustering with an imperative application in finance.

\subsubsection*{Code Availability:} All code necessary to reproduce the experiments and generate the plots are available at: \\ \href{https://github.com/supreethmv/Quantum-Asset-Clustering}{https://github.com/supreethmv/Quantum-Asset-Clustering}

\bibliographystyle{splncs04}
\bibliography{references}

\end{document}

%% file: preamble.tex
\usepackage{amsmath}
\usepackage{hyperref}
\usepackage{color}
\usepackage{graphicx}
\usepackage{cite}
\usepackage{amsmath}
\usepackage{amssymb}
\hypersetup{
    colorlinks=true,
    linkcolor=blue,      
    citecolor=green,     
    urlcolor=blue,       
    anchorcolor=blue     
}

\newcommand\blfootnote[1]{%
  \begingroup
  \renewcommand\thefootnote{}\footnote{#1}%
  \addtocounter{footnote}{-1}%
  \endgroup
}

\usepackage{float}